\documentclass{aip-cp}
\usepackage[english]{babel}
\usepackage{rotating}
\usepackage{lscape}
\usepackage{graphicx}
\usepackage{color, multirow,amsfonts}
\begin{document}
\title{Measuring heavy-tailedness of distributions}
\author[aff1]{Pavlina K. Jordanova\corref{cor1}}
\author[aff2]{Monika P. Petkova,}
\eaddress{monikapetevapetkova@abv.bg}
\affil[aff1]{Faculty of Mathematics and Informatics, Shumen University, 115 "Universitetska" str., 9712 Shumen, Bulgaria.}
\affil[aff2]{Faculty of Mathematics and Informatics, Sofia University, 5 "James Bourchier" blvd., 1164 Sofia, Bulgaria.}

\corresp[cor1]{Corresponding author: pavlina\_kj@abv.bg}

\maketitle

\begin{abstract}
Different questions related with analysis of extreme values and outliers arise frequently in practice. To exclude extremal observations and outliers is not a good decision, because they contain important information about the observed distribution. The difficulties with their usage are usually related with the estimation of the tail index in case it exists. There are many measures for the center of the distribution, e.g. mean, mode, median. There are many measures for the variance, asymmetry and kurtosis, but there is no easy characteristic for heavy-tailedness of the observed distribution. Here we propose such a measure, give some examples and explore some of its properties. This allows us to introduce classification of the distributions, with respect to their heavy-tailedness. The idea is to help and navigate practitioners for accurate and easier work in the field of probability distributions.

Using the properties of the defined characteristics some distribution sensitive extremal index estimators are proposed and their properties are partially investigated.
\end{abstract}

\section{INTRODUCTION}

More than 90 years scientists look for appropriate way for handling outliers. \cite{irwin1925criterion}, \cite{mckay1935distribution},   \cite{nair1948distribution} and  \cite{dixon1950analysis, dixon1953processing} consider them mainly with respect to the deviations of the distribution of the maxima of the sample from the one of the maxima of the normal distribution. They discuss the effect of removing outliers and propose some techniques for handling them. Further on some other tests for outliers appear, see e.g. Grubbs' test \cite{grubbs1969procedures}. They still neglects the importance of the extreme values, do not take into account the fact that the standard deviation does not obligatory exists, especially in case of heavy tailed distributions, and compare the observed variable with the appropriate normal one. Recently  \cite{klebanov2016big, klebanov2017outliers, klebanov2016outliers} reminded this topic.
In 1978 Tukey et al. give different definitions for mild and extremal outliers \cite{tukey1977exploratory} and box-plots \cite{mcgill1978variations} via the quartiles of the distribution and the inter-quartile range ($IQR$). Here we make classification of the distributions, with respect to the heaviness of their tails using the theoretical: quartiles $Q_1, Q_2, Q_3$, $IQR$, lower inner fences ($I_L$), lower outer fences ($O_L$), upper inner fences ($I_R$) and upper outer fences ($O_R$).

\begin{figure}
    \includegraphics[scale=.67,draft=false]{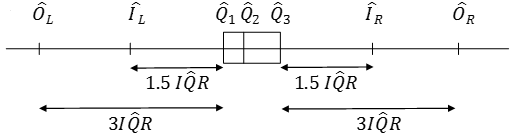}\vspace{-0.3cm}

    \caption{Empirical box-plot, together with the empirical inner and outer fences.}\label{fig:empiricalboxplot}
\end{figure}

Suppose $X_1, X_2, ..., X_n$ are mutually independent observations of a random variable (r.v.) $X$ with cumulative distribution function (c.d.f.) $F(x) = \mathbb{P}(X \leq x)$, probability density function (p.d.f.) $f$ and increasing order statistics $X_{(1, n)} \leq ... \leq X_{(n, n)}$.
There are many different possibilities to define empirical $p$-quantiles, $p \in (0, 1)$. See e.g. \cite{parzen1979nonparametric, hyndman1996sample, langford2006quartiles}. We use the following one $\hat{F}^\leftarrow(p) = X_{([(n+1)p],n)} + \{(n+1)p-[(n+1)p]\}\{X_{([(n+1)p]+1, n)} - X_{([(n+1)p], n)}\}
$, where $[a]$ means the integer part of $a$ and $\frac{1}{n+1} \leq p \leq \frac{n}{n+1}$. 
Let $\hat{Q}_1$, $\hat{Q}_2$, $\hat{Q}_3$ be the empirical quartiles of the observed r.v. and $\hat{IQR} = \hat{Q}_3 - \hat{Q}_1$ be the corresponding empirical IQR.  We use the concepts for empirical: lower inner fences $\hat{I}_L = \hat{Q}_1 - 1.5×\hat{IQR}$, upper inner fences $\hat{I}_R = \hat{Q}_3 + 1.5×\hat{IQR}$, lower outer fences $\hat{O}_L = \hat{Q}_1 - 3×\hat{IQR}$, upper outer fences $\hat{O}_R = \hat{Q}_3 + 3×\hat{IQR}$, mild and extreme outliers, given e.g. in  \cite{devore2015probability, NISTSEMATHECH, watkins2010statistics}.  We call an observation {\bf{mild outlier}} if it is outside the interval $[\hat{Q}_1 - 1.5\hat{IQR}; \hat{Q}_3 + 1.5\hat{IQR}]$ and inside the interval $[\hat{Q}_1 - 3\hat{IQR}; \hat{Q}_3 + 3\hat{IQR}]$. We call an observation {\bf{extreme outlier}} if it is outside the interval $[\hat{Q}_1 - 3\hat{IQR}; \hat{Q}_3 + 3\hat{IQR}]$. See  Figure~\ref{fig:empiricalboxplot} and \cite{devore2015probability}.

Different questions related with analysis of outliers arise frequently in practice. The difficulties with their usage are usually related with the estimation of the tail index in case it exists. Recently the extreme value theory develops techniques for handling them, but it mainly relies on the second order condition (see e.g. \cite{de2007extreme}). It seems to be difficult to be checked, handled and understood from practitioners. Due to luck of information about the distribution outside the range of the data, its tail should be estimated via many characteristics. There are many measures for the center of the distribution, e.g. mean, mode, median. There are measures for the variance, asymmetry and kurtosis, but there is no enough characteristics for measuring heaviness of the tails of the distribution. Here we propose such measures and give some examples. All of them are invariant with respect to shifting of the discussed r.v. This allows us to introduce classification of the distributions, with respect to their heavy-tailedness. Using the outliers we propose a relatively easy techniques to recognize the tail of the distribution and to estimate its index of regular variation in case it exists.  The idea is to help and navigate practitioners for accurate statistical diagnostics and easier work in the field of probability distributions. This approach provides benchmarks only for recognizing the tails of the observed  distribution. For better fit we need to take into account also the specific form of its center.

\section{CLASSIFICATION OF DISTRIBUTIONS WITH RESPECT TO THEIR HEAVY-TAILEDNESS}

\begin{figure}
\includegraphics[scale=.67,draft=false]{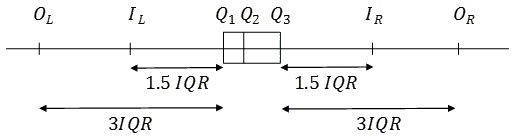}\vspace{-0.3cm}

\caption{Theoretical box-plot, together with theoretical inner and outer fences.}\label{fig:Fboxplot}
\end{figure}

Following Tukey, under theoretical box-plot of a given c.d.f. $F$ we understand the one on Figure~\ref{fig:Fboxplot}.  One of the possibilities to make a tentative fitting of the observed distribution is to compare its empirical box-plot with the theoretical box-plot of the tested distribution. However this approach is not robust, especially for small samples. See e.g. \cite{devore2015probability}. The presence of outliers in a sample of independent observations strongly depends not only of the distributional type, but also from the sample size. Therefore we classify the distributions with respect to their probabilities to have mild or extreme outliers. First of all let us mention that all numerical characteristics that we introduce are invariant with respect to shifting of the r.v.

\subsection{Classification of the distributions with respect to heaviness of their left tails}

{\bf{Definition 1.}} We call a r.v. $X$ and its c.d.f. $F$, {\it $p_{mL}(X)$-mild-heavy left-tailed} if
$$P(Q_{1}(F) - 3IQR(F) < X \leq Q_{1}(F) - 1.5IQR(F)) = p_{mL}(X).$$

Having in mind this definition we introduce classification of the distributions with respect to their mild left tail.

\begin{figure}
\begin{minipage}[t]{0.5\linewidth}
\centerline{%
    \includegraphics
    [width=.78\textwidth]{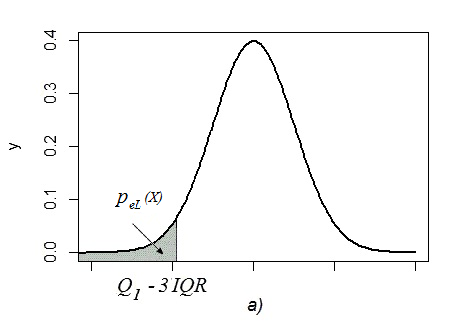}}
\end{minipage}
\begin{minipage}[t]{0.5\linewidth}
\centerline{%
    \includegraphics
[width=.78\textwidth]{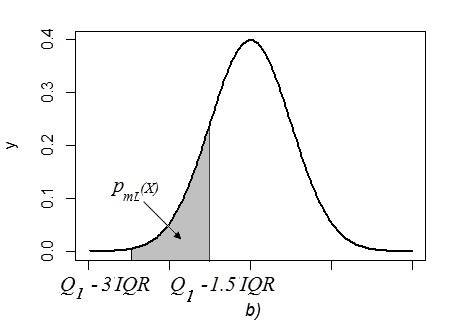}}
\caption{Relation between the plot of the p.d.f. of a r.v. $X$ with c.d.f. $F$, $p_{mL}(X)$ and $p_{eL}(X)$. \label{fig:Fig3}}
\end{minipage}
\end{figure}

{\bf{Definition 2.}} A r.v. $X$ and a r.v. $Y$ {\it belong to one and the same $p_{mL}$-mild-heavy left-tailed class} if $p_{mL}(X) = p_{mL}(Y)$. See Figure~\ref{fig:Fig3}, b). {\it A r.v. $X$ has lighter mild-heavy left tail than a r.v. $Y$} if $p_{mL}(X) < p_{mL}(Y)$.

Let us note that $p_{mL}(X) = p_{mL}(Y)$ does not mean neither that the $X$ and $Y$ belong to one and the same distributional type, nor that they have one and the same mean or variance. But if $X = Y$  in distribution then $p_{mL}(X) = p_{mL}(Y)$.

The $p_{mL}$ characteristic is invariant with respect to shifting. More precisely, for all $c_1 \in \mathbb{R}$
$p_{mL}(c_1 + X) = p_{mL}(X).$

Table~\ref{tab:1} presents a small part of the this classification, where $c_m = \left(\log_{1-\frac{log\,3}{log\,4}}\frac{3}{5}\right)^{-1} \approx 3.08$, $c_e = \left(\log_{1-\frac{log\,3}{log\,4}}\frac{3}{4}\right)^{-1} \approx 5.47$. The fact that $p_{mL}$ characteristic of all normal distributions is approximately $0.0035$ in practice means that if we observe such a r.v. we should expect 3 or 4 mild left outliers to appear in a sample of 1000 observations. Analogously we should expect to have around 34 or 35 mild left outliers in a sample of 10000 observations and so on. All negative exponential distributions have approximately $0.0203$-mild-heavy left tail. So, if we observe 100 independent realizations of exponentially distributed r.v. we should expect to have 2 mild  left outliers.

\begin{table}
\centering
\caption{Classification of some of the distributions with respect to their mild-heavy left-taildeness.   \label{tab:1}}

\begin{tabular}{|c|c|}
  \hline
    Distribution & $p_{mL}$  \\
   \hline
   $U(a, b), a < b; Gamma(\alpha, \beta); Pareto(\alpha, \delta); Frechet(\alpha), 0 < \alpha <  c_m$      & $0 $ \\
  $Frechet(\alpha), \alpha \in  (c_m, c_e]$  & $exp\{-\left(2.5 log^{-1/\alpha}(4)-1.5log^{-1/\alpha}\frac{4}{3} \right)^{-\alpha}\} \approx 0$\\
    $Frechet(\alpha), \alpha >  c_e$  & $exp\{-\left(2.5 log^{-1/\alpha}(4)-1.5log^{-1/\alpha}\frac{4}{3} \right)^{-\alpha}\}-$\\
      & $-exp\{-\left(4 log^{-1/\alpha}4-3log^{-1/\alpha}\frac{4}{3} \right)^{-\alpha}\} \approx 0$\\
       $Gumbel$ & $\approx  0.00000043$\\
    $N(\mu, \sigma^2)$   & $\approx 0.0035$ \\
     $Weibull^-(\alpha)$ & $exp\{-(2.5log^{1/\alpha}4 -1.5 log^{1/\alpha}\frac{4}{3})^{\alpha} \}-$\\
  & \hfill$-exp\{-(4 log^{1/\alpha}4 - 3 log^{1/\alpha}\frac{4}{3})^{\alpha} \}$\\
   $Weibull^-(2)$  & $\approx 0.0102$\\
      $t(2)$  & $\approx 0.0266$  \\
  $t(1)$ & $\approx 0.0328$  \\
    $-Exp(\lambda)$  & $\approx 0.0339$  \\
   $Weibull^-(1)$  & $\approx 0.0389$\\
  $Weibull^-(0.5)$  & $\approx 0.0495$\\
          \hline
\end{tabular}
\end{table}
What about more extreme left outliers? See Figure~\ref{fig:Fig3}, a).

{\bf{Definition 3.}} We call a r.v. $X$ and its c.d.f. $F$, {\it $p_{eL}(X)$-extremely heavy left-tailed} if
$$P(X < Q_{1}(F) - 3IQR(F)) = p_{eL}(X).$$

{\bf{Definition 4.}} We say that {\it a r.v. $X$ and a r.v. $Y$   belong to one and the same $p_{eL}$-extremely heavy left-tailed class} if $p_{eL}(X) = p_{eL}(Y)$. Analogously, we say that {\it a r.v. $X$ has lighter extremely heavy left tail than a r.v. $Y$} if $p_{eL}(X) < p_{eL}(Y)$.

Table~\ref{tab:2} presents some examples of classification of distributions with respect to their extremely heavy left tails. In order to explain the results let us consider again the normal distribution. The value $p_{eL} \approx 0.0000012$ means that in case we have independent observations on such a r.v. we should expect to have 1 or 2 left extreme outliers in a sample of $10^6$ observations. Analogously we should expect to have approximately 12 left extreme outliers in a sample of $10^7$ observations and so on.

\begin{table}
\centering
\caption{Classification of some of the distributions with respect to their extremely heavy left-tailedness. \label{tab:2}}

\begin{tabular}{|c|c|}
  \hline
   Distribution & $p_{eL} = F(O_L)$ \\
  \hline
   U(a, b); Gamma$(\alpha, \lambda)$;  Pareto($\alpha$, $\delta$); Frechet($\alpha$), $0 < \alpha <  c_e$    &  0 \\
       Frechet($\alpha$), $\alpha \geq  c_e$; Gumbel     & $\approx 0$\\
  N($\mu$, $\sigma^2$)      &  $\approx 0.0000012$  \\
      $Weibull^-(\alpha)$  & $exp\{-(4 log^{1/\alpha}4 - 3 log^{1/\alpha}\frac{4}{3})^{\alpha}$\\
    $Weibull^-(2)$  & $\approx 0.0000668$\\
      $-Exp(\lambda)$  & $\approx 0.0093$  \\
    $Weibull^-(1)$  & $\approx 0.0093$\\
        $t(2)$  & $\approx 0.0146$  \\
      $t(1)$  & $\approx 0.0452$  \\
      $Weibull^-(0.5)$  & $\approx 0.0654$\\

      \hline
\end{tabular}
\end{table}

{\it Note:} 1. $p_{mL}(X) < p_{mL}(Y)$ is not equivalent to $p_{eL}(X) < p_{eL}(Y)$.

2. If $p_{eL}(X) = p_{eL}(Y)$ or  $p_{mL}(X) = p_{mL}(Y)$,  this does not obligatory mean that $X$ and $Y$ coincide in distribution.

3. $p_{eL}(c_1 + X) = p_{eL}(X)$, for all $c_1 \in \mathbb{R}$.

\subsection{Classification of the distributions with respect to heaviness of their right tails}

\begin{figure}
\begin{minipage}[t]{0.5\linewidth}
\centerline{%
    \includegraphics
    [width=.78\textwidth]{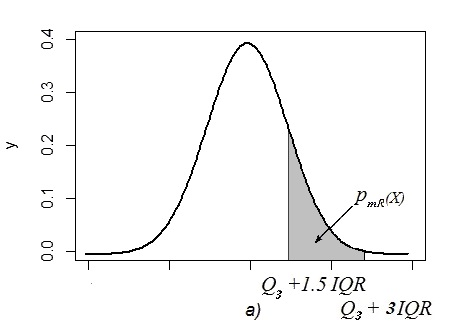}}
\end{minipage}
\begin{minipage}[t]{0.5\linewidth}
\centerline{%
    \includegraphics
[width=.78\textwidth]{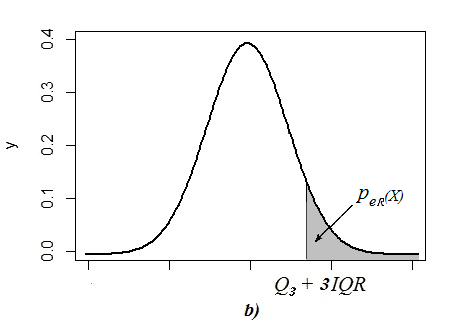}}
\caption{Relation between the plot of the p.d.f. of a r.v. $X$ with c.d.f. $F$, $p_{mR}(X)$ and $p_{eR}(X)$. \label{fig:Fig4}}
\end{minipage}
\end{figure}

Analogously to the previous subsection we can work with the right tails. See Figure~\ref{fig:Fig4}, a) and b).

{\bf{Definition 5.}}  We call a r.v. $X$ and its c.d.f. $F$, {\it $p_{mR}(X)$-mild-heavy right-tailed} if
$$P(Q_{3}(F) + 1.5IQR(F) < X \leq Q_{3}(F) + 3IQR(F)) = p_{mR}(X).$$


{\bf{Definition 6.}} We say that {\it  a r.v. $X$ and a r.v. $Y$ belong to one and the same $p_{mR}$-mild-heavy right-tailed class} if $p_{mR}(X) = p_{mR}(Y)$. {\it  A r.v. $X$ has lighter mild-heavy right tail than a r.v. $Y$} if $p_{mR}(X) < p_{mR}(Y)$.

{\bf{Definition 7.}}  We call a r.v. $X$ and its c.d.f. $F$, {\it $p_{eR}(X)$-extremely heavy right-tailed} if $$P(X > Q_{3}(F) + 3IQR(F)) = p_{eR}(X).$$

{\bf{Definition 8.}} {\it A r.v. $X$ and a r.v. $Y$  belong to one and the same $p_{eR}$-extremely heavy right-tailed  class} if $p_{eR}(X) = p_{eR}(Y)$.  We say that {\it  a r.v. $X$ has lighter extreme right tail than a r.v. $Y$} if $p_{eR}(X) < p_{eR}(Y)$.

The properties of these characteristics are analogous to the corresponding one of the left tails.

 Some examples are given in Table~\ref{tab:3}. Again we observe that $p_{mR}(X) < p_{mR}(Y)$ is not equivalent to $p_{eR}(X) < p_{eR}(Y)$. The analysis is analogous to those made above for the left tails. It is well known that if we consider only a fixed distribution with regularly varying tail, the bigger the value of $\alpha$ the lighter the corresponding tail of the distribution is. However when we consider the extremely heavy tails, which one of Pareto or Frechet distribution has heavier right tail depends on their parameters. If $X \sim Pareto(2, 1)$ and  $Y \sim Frechet(\alpha)$, $\alpha \geq 1$ then $X$ has heavier right tail than $Y$, but $Frechet(0.5)$  has heavier extremal right tail than $Pareto(0.5, 1)$.

\begin{table}
\centering
\caption{Classification of some of the distributions with respect to heaviness of their right tails. \label{tab:3}}
\begin{tabular}{|c|c|c|}
  \hline
  Distribution  & $p_{mR}$ & $p_{eR} = \bar{F}_X(O_R)$ \\
   \hline
  $U(a, b), a < b, Weibull^-(\alpha),\, a, b \in \mathbb{R} $      & $0$ &  $0$ \\
 $ N(\mu, \sigma^2)$ & $\approx 0.0035$  &  $\approx 0.0000012$  \\
 $Gamma(2, \lambda )$, $\lambda  > 0$    &  $\approx 0.0011$ & $\approx 0.000071$ \\
  $Gumbel$   &  $\approx 0.0243$ & $\approx 0.0026$ \\
   $Exp(\lambda), \, \lambda > 0$   &  $\approx 0.0339$  & $\approx 0.0093$  \\
 $t(2)$  & $\approx 0.0266$  & $\approx 0.0146$ \\
  $Gamma(0.5, \lambda )$, $\lambda  > 0$    &  $\approx \mathbf{0.0502}$ & $\approx 0.0255$ \\
  $Frechet(\alpha)$     &  $1 - exp^{-(2.5\sqrt[\alpha]{3.48} - 1.5\sqrt[\alpha]{0.72})^{-\alpha}}$ & $1 - e^{-(4\sqrt[\alpha]{3.48} - 3\sqrt[\alpha]{0.72})^{-\alpha}}$\\
  $Frechet(2)$       &  $\approx 0.0429$ & $\approx 0.0406$\\
  $t(1)$  & $\approx 0.0328$  & $\approx 0.0452$ \\
   $Pareto(\alpha, \delta)$ & $\frac{\delta^{-\alpha}}{4}(2.5-1.5\sqrt[\alpha]{\frac{1}{3}})^{-\alpha} - p_{eR}$  & $ \frac{\delta^{-\alpha}}{4}(4-3\sqrt[\alpha]{\frac{1}{3}})^{-\alpha}$  \\
  $Pareto(2, 1) $    &  $\approx \mathbf{0.045}$ & $\approx 0.0486$  \\
    $Frechet(1)$       &  $\approx 0.0415$ & $\approx 0.0817$ \\
  $Pareto(1, 1)$     &  $\approx 0.0417$  & $\approx 0.0833$ \\
   $Pareto(0.5, 1)$   &  $\approx 0.0331$ & $\approx 0.1306$ \\
  $Frechet(0.5)$     &  $\approx 0.0323$ & $\approx 0.1360$\\
         \hline
\end{tabular}
\end{table}

Note that if $X \sim Gamma(\alpha, \lambda)$, $\lambda > 0$, then $p_{mL}(X)$, $p_{eL}(X)$, $p_{mR}(X)$ and $p_{eR}(X)$ does not depend on $\lambda$.

\subsection{Classification of the distributions with respect to heaviness of their two-sided tails}

Here, for the seek of completeness, we consider the two-sided heavy-tailedness of the distributions. However in practice it is better to make a more detailed  comparison of the probabilities to have one-sided  left or right, mild or extreme outliers. It gives us more comprehensive picture about the tail behaviour of the observed distribution.

{\bf{Definition 9.}}  We call a r.v. $X$ and its c.d.f. $F$, {\it $p_{m2}(X)$-mild-heavy two-tailed} if
$$P(Q_{1}(F) - 3IQR(F) < X \leq Q_{1}(F) - 1.5IQR(F) \cup Q_{3}(F) + 1.5IQR(F) < X \leq Q_{3}(F) + 3IQR(F)) = p_{m2}(X).$$

{\bf{Definition 10.}} {\it A r.v. $X$ and a r.v. $Y$  belong to one and the same $p_{m2}$-mild-heavy two-tailed class} if $p_{m2}(X) = p_{m2}(Y)$. {\it A r.v. $X$ with c.d.f. $F$ has lighter mild two-tails than a r.v. $Y$}  if $p_{m2}(X) < p_{m2}(Y)$.

{\bf{Definition 11.}} A r.v. $X$ and its c.d.f. $F$ are called {\it $p_{e2}(X)$-extremely heavy two-tailed} if
$$P(X < Q_{1}(F) - 3IQR(F) \cup X > Q_{3}(F) + 3IQR(F)) = p_{e2}(X).$$

 {\bf{Definition 12.}} {\it A r.v. $X$ and a r.v. $Y$  belong to one and the same $p_{e2}$-extremely heavy two-tailed class} if $p_{e2}(X) = p_{e2}(Y)$ and {\it a r.v. $X$ has lighter extreme two-tails than a r.v. $Y$} if $p_{e2}(X) < p_{e2}(Y)$.

 {\it Note:} Again the equalities $p_{m2}(X) = p_{m2}(Y)$ or $p_{e2}(X) = p_{e2}(Y)$, does not obligatory mean that $X \stackrel{\rm d}{=} Y$.

\begin{figure}
\begin{minipage}[t]{0.5\linewidth}
\centerline{%
    \includegraphics
    [width=.78\textwidth]{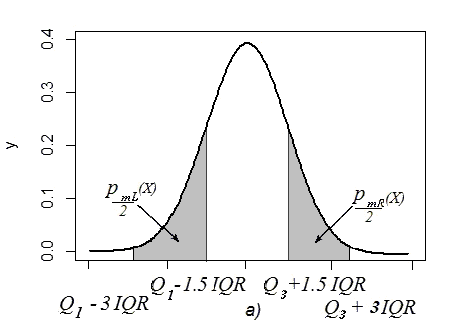}}
\end{minipage}
\begin{minipage}[t]{0.5\linewidth}
\centerline{%
    \includegraphics
[width=.78\textwidth]{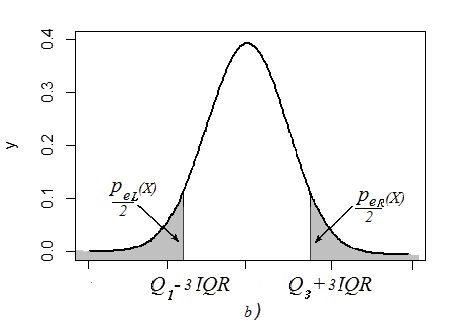}}
\caption{Relation between the plot of the p.d.f. of a r.v. $X$, $p_{m2}(X) = \frac{p_{mL}(X)+p_{mR}(X)}{2}$ and $p_{e2}(X)= \frac{p_{eL}(X)+p_{eR}(X)}{2}$. \label{fig:Fig5}}
\end{minipage}
\end{figure}

 In Table~\ref{tab:6} we have presented the values of $p_{m2}(X)$ and $p_{e2}(X)$ for some probability laws. See Figure~\ref{fig:Fig5}, a) and b).

\begin{table}
\centering
\caption{Classification of some of the distributions with respect to heaviness of their two-sided tails. \label{tab:6}}
\begin{tabular}{|c|c|c|}
  \hline
  Distribution  & $p_{m2}(X)  = p_{mL}(X) + p_{mR}(X)$ & $p_{e2}(X) = F_X(O_L) + \bar{F}_X(O_R)$ \\
   \hline
  U(a, b)      & 0 & 0  \\
  N($\mu$, $\sigma^2$)  &$\approx 0.007 $ & $\approx 0.000002$ \\
  $Gamma(2, \lambda )$, $\lambda  > 0$    &  $\approx 0.0011$ & $\approx 0.000071$ \\
     $Weibull^-(\alpha)$  & $exp\{-(2,5 log^{1/\alpha}4 - 1,5log^{1/\alpha}\frac{4}{3} )^{\alpha}\} - p_1$ & $p_1 = exp\{-(4 log^{1/\alpha}4-3log^{1/\alpha}\frac{4}{3})^\alpha\}$\\
      $Weibull^-(2)$  & $\approx 0.0102$ & $\approx 0.000067$\\
      $Gumbel$   &  $\approx 0.0243$ & $\approx 0.0026$ \\
       $-Exp(\lambda),\, Exp(\lambda), \, \lambda > 0$   &  $\approx 0.0339$  & $\approx 0.0093$  \\
  $Weibull^-(1)$  & $\approx 0.0388$ & $\approx 0.0093$\\
   $Gamma(0.5, \lambda )$, $\lambda  > 0$    &  $\approx \mathbf{0.0501}$ & $\approx 0.0255$ \\
  $t(2)$  & $\approx \mathbf{0.0532}$  & $\approx 0.0293$ \\
   $Frechet(\alpha)$     &  $1 - exp^{-(2.5\sqrt[\alpha]{3.48} - 1.5\sqrt[\alpha]{0.72})^{-\alpha}} - p_3$ & $p_3 = 1 - e^{-(4\sqrt[\alpha]{3.48} - 3\sqrt[\alpha]{0.72})}$\\
   $Frechet(2)$       &  $\approx 0.0429$ & $\approx 0.0406$\\
    $ Pareto(\alpha, \delta)$ & $\frac{\delta^{-\alpha}}{4}(2.5-1.5\sqrt[\alpha]{\frac{1}{3}})^{-\alpha} - p_2$  & $p_2 = \frac{\delta^{-\alpha}}{4}(4-3\sqrt[\alpha]{\frac{1}{3}})^{-\alpha}$  \\
  $Pareto(2, 1) $    &  $\approx 0.045$ & $\approx 0.0486$  \\
       $Weibull^-(0.5)$  & $\approx 0.0495$  & $\approx 0.0654$\\
  $Frechet(1)$       &  $\approx 0.0415$ & $\approx 0.0817$ \\
  $Pareto(1, 1)$     &  $\approx 0.0417$  & $\approx 0.0833$ \\
  $t(1)$  & $\approx \mathbf{0.0656}$  & $\approx 0.0903$ \\
  $Pareto(0.5, 1)$   &  $\approx 0.0331$ & $\approx 0.1306$ \\
    $Frechet(0.5)$     &  $\approx 0.0323$ & $\approx 0.1360$\\
        \hline
\end{tabular}
\end{table}

\subsection{Algorithm for applications}
Considering the outliers in a sample and comparing their relative frequencies with $p_{mL}$, $p_{eL}$, $p_{mR}$ and $p_{eR}$ we are able to make a better modelling of the tails of the distribution of the observed r.v. The algorithm is the following:
\begin{enumerate}
   \item Determine $\hat{Q}_1$, $\hat{Q}_2$, $\hat{Q}_3$, $I\hat{Q}R$, $\hat{I}_L$, $\hat{O}_L$, $\hat{I}_R$, $\hat{O}_R$ and compare the empirical box-plot with the theoretical box-plot of the chosen distributions.
  \item Determine the relative frequencies of the left and right, mild and extreme outliers.
  \item Make confidence intervals, based on the relative frequencies of mild or extreme outliers. Compare these relative frequencies with $p_{mL}$ and $p_{mR}$ in the list of distributions and chose appropriate classes of distributions for modelling the probability law of the observed r.v.
  \item Make confidence intervals, based only on the relative frequencies of extreme outliers. Compare these relative frequencies with $p_{eL}$ and $p_{eR}$ in the list of distributions chosen in 3. and find the most appropriate distributional types for modeling the observed r.v.
  \item Estimate the parameters of the chosen distributions.
  \item Use some goodness of fit test to chose the best model.
 \end{enumerate}

\section{FIVE NEW ESTIMATORS OF THE EXTREMAL INDEX. EMPIRICAL STUDY.}

In this section we suppose that $\widehat{Q}_1 > 0,\widehat{Q}_1 \not= 1$ and at least one of the following two conditions hold: $\widehat{Q}_1\not = \widehat{Q}_3$ or $\widehat{O}_R > 1$. We propose to model the observed r.v. with appropriate distribution with regularly varying tail, i.e. such that
$\lim_{y \to \infty} \frac{1 - F(xy)}{1 - F(y)} = x^{-\alpha}$
 and present five distribution sensitive estimators of the parameter $\alpha$. The relative frequency $\hat{p}_{eR}$ of the right extreme outliers in the sample is a strongly consistent and unbiased estimator of $p_{eR}$. The right outer fence $\hat{O}_R$ is an asymptotically consistent estimator of the theoretical $O_R$.

 The following two estimators have very fast rate of convergence in case when the observed  r.v. is $Pareto(\alpha)$ distributed. See the empirical study and  Table~\ref{tab:estimators}.
$$\hat{\alpha}_{Par,n} =  -\frac{log\, \widehat{p}_{eR}}{log\,\widehat{O}_R}, \quad \hat{\alpha}_{Q, Par,n} = \frac{log(3)}{log \widehat{Q}_3 - log \widehat{Q}_1}.$$
They have approximately the same properties as the Hill and the t-Hill estimators.
\begin{landscape}
\begin{table}
\centering
\caption{Empirical results.   \label{tab:estimators}}

\begin{tabular}{|c|c|c|c|c|c|c|c|c|c|c|c|c|c|}
  \hline
  \multicolumn{2}{|c|}{Distribution}  & \multirow{2}{*}{$n$}  &\multicolumn{2}{c|}{$\hat{\alpha}_{Par,n}$}    & \multicolumn{2}{c|}{$\hat{\alpha}_{Q, Par,n}$}   & \multicolumn{2}{c|}{$\hat{\alpha}_{Frech, n}$}    & \multicolumn{2}{c|}{$\hat{\alpha}_{Q, Frech, n}$}   & \multicolumn{2}{c|}{$\hat{\alpha}_{Q, HillH, n}$}  & \multirow{2}{*}{The best}\\
  \cline{4-13}
 \multicolumn{2}{|c|}{of $\xi$.}   &   & Mean & St. Dev. & Mean & St. Dev. & Mean & St. Dev. & Mean & St. Dev. & Mean & St. Dev.& estimator\\ \hline
\multirow{12}{*}{$Pareto(\alpha)$} & \multirow{4}{*}{$\alpha = 0.5$}  & $30$ & 0.5463 & 0.1356&  {\bf 0.5116} & 0.1442 & 0.531 &  0.138 & 0.7323 &0.2064  & 1.2437 &  29.8589 & $\hat{\alpha}_{Q, Par,n}$\\
&  & $100$ &0.5119& 0.0669 & {\bf 0.5035} & 0.0756 & 0.4954 & 0.0683 &0.7207 &0.1083 & 0.3904 & 74.1974& $\hat{\alpha}_{Q, Par,n}$\\
  &  & $1000$ & 0.5015 & 0.0204 & {\bf 0.5005} & 0.0238 & 0.4846 & 0.0208 & 0.7164 & 0.034& 1.7906 &  0.3105& $\hat{\alpha}_{Q, Par,n}$\\
   &  & $10000$ & {\bf 0.5001} & 0.0063 &  0.5002 & 0.0074 & 0.4831 & 0.0063 & 0.716 &0.0106  & 1.761&  0.0922&$\hat{\alpha}_{Par,n}$\\
  \cline{2-14}
  & \multirow{4}{*}{$\alpha = 1$} & $30$ & 1.0671 &0.2528 & {\bf 1.026} & 0.2869 & 1.05 & 0.2568 & 1.4685 & 0.4106 & -2.1133 & 17.9816 & $\hat{\alpha}_{Q, Par,n}$ \\
  &  & $100$ & 1.0312& 0.1486 & {\bf 1.0092} & 0.1516 & 1.0141 & 0.1518 & 1.4446 & 0.2171 & -3.0747 &9.8974 & $\hat{\alpha}_{Q, Par,n}$\\
 &   & $1000$ & 1.0022 & 0.0422 & {\bf 1.0006} & 0.0469 & 0.9848 & 0.0432 &  1.4322& 0.0671&-2.3396& 0.2569&$\hat{\alpha}_{Q, Par,n}$\\
 &  & $10000$ & 1.0004 & 0.0132 & {\bf 1} & 0.0146& 0.983& 0.0136& 1.4314&0.021& -2.324& 0.081& $\hat{\alpha}_{Q, Par,n}$\\
  \cline{2-14}
   & \multirow{4}{*}{$\alpha = 2$} & $30$ & {\bf1.9996} & 0.4313 & 2.0514 & 0.566 & 1.9789 & 0.4351 & 2.9364 & 0.8102 & -1.134 &  0.2572& $\hat{\alpha}_{Par,n}$\\
  &  & $100$ &  2.0772 & 0.3332 & {\bf2.0185} & 0.3021 & 2.0607 & 0.338 & 2.8893 & 0.4325 & -1.0865 & 0.0938 & $\hat{\alpha}_{Q, Par,n}$\\
&   & $1000$ & 2.0083 & 0.0937 & {\bf2.0021} & 0.094 & 1.9919 & 0.0953 & 2.8658 & 0.1346 &  -1.0766& 0.0273& $\hat{\alpha}_{Q, Par,n}$\\
&   & $10000$ & {\bf2} & 0.0292 & 1.9998 & 0.03 & 1.9836 & 0.0297 & 2.8624 & 0.043 &-1.0737 &0.0087& $\hat{\alpha}_{Par,n}$\\
  \hline
 \multirow{12}{*}{$Frechet(\alpha)$} & \multirow{4}{*}{$\alpha = 0.5$}  & $30$ & 0.5738 &  0.1549& 0.354 & 0.0826 & 0.5568 & 0.1564 & {\bf0.5067} & 0.1182 & 0.9214 & 9.8233& $\hat{\alpha}_{Q, Frech, n}$\\
 &  & $100$ & 0.5335 & 0.0738 & 0.3508 & 0.0442 & 0.5151 & 0.0747 & {\bf0.5021} & 0.0632 & 0.7301 &0.2186& $\hat{\alpha}_{Q, Frech, n}$\\
  &  & $1000$ & 0.5201 & 0.0222 & 0.3494 & 0.0139 & 0.5014 & 0.0225 & {\bf0.5001} & 0.0199 & 0.7012& 0.0564 & $\hat{\alpha}_{Q, Frech, n}$\\
  &   & $10000$ & 0.519 & 0.007 & 0.3494 & 0.0044 & 0.5002 & 0.007 & {\bf0.5001} & 0.0062 & 0.6989&  0.0175 &$\hat{\alpha}_{Q, Frech, n}$\\
  \cline{2-14}
  & \multirow{4}{*}{$\alpha = 1$} & $30$ & 1.0825 & 0.2599 & 0.7066 & 0.1648 & 1.0654 & 0.2637 & {\bf 1.0115} &  0.2358 &  3.2049 & 324.4615 & $\hat{\alpha}_{Q, Frech, n}$\\
  &  & $100$ & 1.0506 & 0.1534 & 0.7023 & 0.0892 & 1.0337 & 0.1565 & {\bf1.0053} & 0.1277 & 5.5908 & 707.2482 &$\hat{\alpha}_{Q, Frech, n}$\\
  &   & $1000$ &  1.0209 & 0.0445 & 0.6989 &  0.0276 & 1.0038 & 0.0454 & {\bf1.0004 } &  0.0396 & -15.003& 1031.205 & $\hat{\alpha}_{Q, Frech, n}$\\
  &   & $10000$ &  1.0174 & 0.014 & 0.6987 & 0.0088 & 1.0003 & 0.0143 & {\bf1.0001} & 0.0126 & 287.2494 & 14959.87& $\hat{\alpha}_{Q, Frech, n}$\\
  \cline{2-14}
   & \multirow{4}{*}{$\alpha = 2$} & $30$ & 1.9469 & 0.3957 & 1.4131 & 0.3275 & 1.9289 & 0.3988 & {\bf2.0227} & 0.4688 & -1.5699 & 1.755& $\hat{\alpha}_{Q, Frech, n}$\\
   &  & $100$ & 2.094 & 0.3418 & 1.4031 & 0.178 & 2.081 & 0.3461 & {\bf2.0084} &  0.2548 & -1.4384& 0.1996 &$\hat{\alpha}_{Q, Frech, n}$\\
  &   & $1000$ & 2.0202 & 0.0978 &1.3975  & 0.0554 & 2.0072 & 0.0993 & {\bf2.0004} & 0.0794 & -1.4013 & 0.0553 & $\hat{\alpha}_{Q, Frech, n}$\\
  &   & $10000$ & 2.0144 & 0.0309 & 1.3974 & 0.0175 & 2.0014 & 0.0313 & {\bf2.0002} & 0.025 & -1.3975 & 0.0176& $\hat{\alpha}_{Q, Frech, n}$\\
  \hline
\multirow{12}{*}{$Hill-$} & \multirow{4}{*}{$\alpha = 0.5$}  & $30$ &{\bf0.4692}  & 0.1223& 0.2921 & 0.0675& 0.4534& 0.1234 & 0.4181 &0.0966  &  2.164 &  0.7827& $\hat{\alpha}_{Par,n}$\\
&  & $100$ &{\bf0.4393}  & 0.0591& 0.2925& 0.0367& 0.4221& 0.0598&0.4186 &0.0525& 0.5647& 0.9664& $\hat{\alpha}_{Par,n}$\\
  &  & $1000$ & 0.4286 & 0.0178& 0.2915& 0.0115& 0.411& 0.0179& 0.4173& 0.0165& {\bf 0.5012} & 0.0342 & $\hat{\alpha}_{Q, HillH, n}$\\
  &   & $10000$ & 0.4278 & 0.0057 & 0.2915 & 0.0037 & 0.4102 & 0.0057 & 0.4172 & 0.0052 &  {\bf 0.5001} & 0.0107 & $\hat{\alpha}_{Q, HillH, n}$\\
  \cline{2-14}
  & \multirow{4}{*}{$\alpha = 1$} & $30$ & {\bf0.8051} & 0.205 & 0.4134 & 0.0926 & 0.788 & 0.2077 & 0.5917 & 0.1325 &  2.6214 &  187.5964& $\hat{\alpha}_{Par,n}$\\
  &  & $100$ & 0.7581 & 0.1056& 0.4127 & 0.0505& 0.7403 & 0.1075 & 0.5908 & 0.0723 & {\bf1.0795}& 0.4421& $\hat{\alpha}_{Q, HillH, n}$\\
 &   & $1000$ & 0.7377 & 0.0315 & 0.4115 & 0.0158 & 0.7194 &  0.0321 & 0.589 & 0.0227 & {\bf1.0055} & 0.0948 & $\hat{\alpha}_{Q, HillH, n}$\\
 &   & $10000$ & 0.7355 & 0.0098 & 0.4113 & 0.005 & 0.7172 & 0.01 & 0.5887 & 0.0072 & {\bf1.0008} & 0.0908& $\hat{\alpha}_{Q, HillH, n}$\\
  \cline{2-14}
  $Horror(\alpha)$  & \multirow{4}{*}{$\alpha = 2$} & $30$ & 1.2185 & 0.283 & 0.5198 & 0.1143 & 1.2019 & 0.2859 & 0.744 & 0.1636 & {\bf1.543} &69.0628& $\hat{\alpha}_{Q, HillH, n}$\\
  &  & $100$ &1.1965  &0.1785  & 0.519 & 0.0631 & 1.1812 & 0.1816 & 0.7429 & 0.0903 & {\bf2.8577} & 28.6543&$\hat{\alpha}_{Q, HillH, n}$\\
&   & $1000$ & 1.1554 & 0.05 & 0.5178 & 0.0196 &  1.1399 & 0.051 & 0.7411 & 0.0281 & {\bf2.0379} &0.3168 & $\hat{\alpha}_{Q, HillH, n}$\\
&   & $10000$ & 1.1529 & 0.0159 & 0.5177 & 0.0062 & 1.1374 & 0.0162 & 0.741 & 0.0089 & {\bf2.0044}& 0.0932& $\hat{\alpha}_{Q, HillH, n}$\\
  \hline
\end{tabular}
\end{table}
\end{landscape}

The second group of two estimators
$$\hat{\alpha}_{Frech, n} =  -\frac{log(-log\, \widehat{p}_{eR})}{log\,\widehat{O}_R},\quad \hat{\alpha}_{Q, Frech, n} = -\frac{log(log(4)) - log(log(4/3))}{log \widehat{Q}_3 - log \widehat{Q}_1}$$
is better in cases when the observed r.v. is a $Frechet(\alpha)$ distributed. We should mention that in both of these cases, it is well known that for estimating the parameter of the Pareto distribution, the Hill estimator (see e.g. \cite{hill1975simple}) is the best estimator. With respect to the robustness their behaviour is comparable with  the one of the t-Hill estimator (see \cite{jordanova2012weak, fabian2009ifas}).

The last estimator is the most appropriate in case the observed r.v. has Hill-Horror distribution
$$F^\leftarrow(p) = (1 - p)^{-1/\alpha}(-log\,(1 - p)), p \in (0, 1), \quad (see \cite{EMK}).$$
This estimator is defined by
$$\hat{\alpha}_{Q, HillH, n} = \frac{log(3)}{log \widehat{Q}_3 +log(log(4/3)) - log \widehat{Q}_1 - log(log(4))}.$$

Let us make a brief empirical investigation of these estimators. For different but fixed $n = 30, 10^2, 10^3, 10^4$, we have made $m = 10^4$ samples with sample size $n$, of observation on one and the same r.v. Within these $m = 10^4$ samples the type and the parameters are one and the same, but in general the types change between $Pareto(\alpha)$, $Frechet(\alpha)$  or $Hill-Horror(\alpha)$ distribution for different $\alpha$. Then we have calculated $10^4$ values of $\hat{\alpha}_{Par,n}$, $\hat{\alpha}_{Q, Par,n}$, $\hat{\alpha}_{Frech, n}$, $\hat{\alpha}_{Q, Frech, n}$, $\hat{\alpha}_{Q, HillH, n}$ and finally  we have calculated the corresponding means and standard deviations. The results are given in Table~\ref{tab:estimators}. The best estimator in any particular case is the one that takes into account the type of the observed r.v. Therefore the choice of the distribution is the most important step for the estimation of the index of regular variation.

Although we have found good estimators for the regularly varying index when the observed distribution is almost regularly varying. The most dangerous case,  is again the Hill-Horror distributed one. The question about estimation of $\alpha$ for small samples of such data, e.g. when $n \leq 30$ is still open. In this cases, however it seems to be not realistic to find a good estimator of the tail index, because due to the slow rate of convergence, with very high probability, the sample does not contain enough information about the tail of the distribution.

\section{ACKNOWLEDGMENTS}
The authors was partially supported by the Project RD-08-96/06.02.2017 from the Scientific Research Fund in
University of Shumen and grant No 80-10-146/21.04.2017 of Sofia University, Bulgaria.


\nocite{*}
\bibliographystyle{aipnum-cp}%

\begin{thebibliography}{22}%
\makeatletter
\providecommand \@ifxundefined [1]{%
 \@ifx{#1\undefined}
}%
\providecommand \@ifnum [1]{%
 \ifnum #1\expandafter \@firstoftwo
 \else \expandafter \@secondoftwo
 \fi
}%
\providecommand \@ifx [1]{%
 \ifx #1\expandafter \@firstoftwo
 \else \expandafter \@secondoftwo
 \fi
}%
\providecommand \natexlab [1]{#1}%
\providecommand \enquote  [1]{``#1''}%
\providecommand \bibnamefont  [1]{#1}%
\providecommand \bibfnamefont [1]{#1}%
\providecommand \citenamefont [1]{#1}%
\providecommand \href@noop [0]{\@secondoftwo}%
\providecommand \href [0]{\begingroup \@sanitize@url \@href}%
\providecommand \@href[1]{\@@startlink{#1}\@@href}%
\providecommand \@@href[1]{\endgroup#1\@@endlink}%
\providecommand \@sanitize@url [0]{\catcode `\$12\catcode `\&12\catcode
  `\#12\catcode `\^12\catcode `\_12\catcode `\%12\relax}%
\providecommand \@@startlink[1]{}%
\providecommand \@@endlink[0]{}%
\providecommand \url  [0]{\begingroup\@sanitize@url \@url }%
\providecommand \@url [1]{\endgroup\@href {#1}{\urlprefix }}%
\providecommand \urlprefix  [0]{URL }%
\providecommand \Eprint [0]{\href }%
\providecommand \doibase [0]{http://dx.doi.org/}%
\providecommand \selectlanguage [0]{\@gobble}%
\providecommand \bibinfo  [0]{\@secondoftwo}%
\providecommand \bibfield  [0]{\@secondoftwo}%
\providecommand \translation [1]{[#1]}%
\providecommand \BibitemOpen [0]{}%
\providecommand \bibitemStop [0]{}%
\providecommand \bibitemNoStop [0]{.\EOS\space}%
\providecommand \EOS [0]{\spacefactor3000\relax}%
\providecommand \BibitemShut  [1]{\csname bibitem#1\endcsname}%
\let\auto@bib@innerbib\@empty
\bibitem [{\citenamefont {Irwin}(1925)}]{irwin1925criterion}%
  \BibitemOpen
  \bibfield  {author} {\bibinfo {author} {\bibfnamefont {J.}~\bibnamefont
  {Irwin}},\ }\href@noop {} {\bibfield  {journal} {\bibinfo  {journal}
  {Biometrika}\ }\unskip\ \bibinfo {pages} {238--250} (\bibinfo {year}
  {1925})}\BibitemShut {NoStop}%
\bibitem [{\citenamefont {McKay}(1935)}]{mckay1935distribution}%
  \BibitemOpen
  \bibfield  {author} {\bibinfo {author} {\bibfnamefont {A.}~\bibnamefont
  {McKay}},\ }\href@noop {} {\bibfield  {journal} {\bibinfo  {journal}
  {Biometrika}\ }\textbf {\bibinfo {volume} {27}},\ \unskip\ \bibinfo {pages}
  {466--471} (\bibinfo {year} {1935})}\BibitemShut {NoStop}%
\bibitem [{\citenamefont {Nair}(1948)}]{nair1948distribution}%
  \BibitemOpen
  \bibfield  {author} {\bibinfo {author} {\bibfnamefont {K.}~\bibnamefont
  {Nair}},\ }\href@noop {} {\bibfield  {journal} {\bibinfo  {journal}
  {Biometrika}\ }\textbf {\bibinfo {volume} {35}},\ \unskip\ \bibinfo {pages}
  {118--144} (\bibinfo {year} {1948})}\BibitemShut {NoStop}%
\bibitem [{\citenamefont {Dixon}(1950)}]{dixon1950analysis}%
  \BibitemOpen
  \bibfield  {author} {\bibinfo {author} {\bibfnamefont {W.~J.}\ \bibnamefont
  {Dixon}},\ }\href@noop {} {\bibfield  {journal} {\bibinfo  {journal} {The
  Annals of Mathematical Statistics}\ }\textbf {\bibinfo {volume} {21}},\
  \unskip\ \bibinfo {pages} {488--506} (\bibinfo {year} {1950})}\BibitemShut
  {NoStop}%
\bibitem [{\citenamefont {Dixon}(1953)}]{dixon1953processing}%
  \BibitemOpen
  \bibfield  {author} {\bibinfo {author} {\bibfnamefont {W.}~\bibnamefont
  {Dixon}},\ }\href@noop {} {\bibfield  {journal} {\bibinfo  {journal}
  {Biometrics}\ }\textbf {\bibinfo {volume} {9}},\ \unskip\ \bibinfo {pages}
  {74--89} (\bibinfo {year} {1953})}\BibitemShut {NoStop}%
\bibitem [{\citenamefont {Grubbs}(1969)}]{grubbs1969procedures}%
  \BibitemOpen
  \bibfield  {author} {\bibinfo {author} {\bibfnamefont {F.~E.}\ \bibnamefont
  {Grubbs}},\ }\href@noop {} {\bibfield  {journal} {\bibinfo  {journal}
  {Technometrics}\ }\textbf {\bibinfo {volume} {11}},\ \unskip\ \bibinfo
  {pages} {1--21} (\bibinfo {year} {1969})}\BibitemShut {NoStop}%
\bibitem [{\citenamefont {Klebanov}(2016)}]{klebanov2016big}%
  \BibitemOpen
  \bibfield  {author} {\bibinfo {author} {\bibfnamefont {L.~B.}\ \bibnamefont
  {Klebanov}},\ }\href@noop {} {\bibfield  {journal} {\bibinfo  {journal}
  {arXiv preprint arXiv:1611.05410}\ } (\bibinfo {year} {2016})}\BibitemShut
  {NoStop}%
\bibitem [{\citenamefont {Klebanov}\ \emph {et~al.}(2017)\citenamefont
  {Klebanov}, \citenamefont {Antoch}, \citenamefont {Karlova},\ and\
  \citenamefont {Kakosyan}}]{klebanov2017outliers}%
  \BibitemOpen
  \bibfield  {author} {\bibinfo {author} {\bibfnamefont {L.~B.}\ \bibnamefont
  {Klebanov}}, \bibinfo {author} {\bibfnamefont {J.}~\bibnamefont {Antoch}},
  \bibinfo {author} {\bibfnamefont {A.}~\bibnamefont {Karlova}}, \ and\
  \bibinfo {author} {\bibfnamefont {A.~V.}\ \bibnamefont {Kakosyan}},\
  }\href@noop {} {\bibfield  {journal} {\bibinfo  {journal} {arXiv preprint
  arXiv:1701.06642}\ } (\bibinfo {year} {2017})}\BibitemShut {NoStop}%
\bibitem [{\citenamefont {Klebanov}, \citenamefont {Kakosyan},\ and\
  \citenamefont {Karlova}(2016)}]{klebanov2016outliers}%
  \BibitemOpen
  \bibfield  {author} {\bibinfo {author} {\bibfnamefont {L.~B.}\ \bibnamefont
  {Klebanov}}, \bibinfo {author} {\bibfnamefont {A.~V.}\ \bibnamefont
  {Kakosyan}}, \ and\ \bibinfo {author} {\bibfnamefont {A.}~\bibnamefont
  {Karlova}},\ }\href@noop {} {\bibfield  {journal} {\bibinfo  {journal} {arXiv
  preprint arXiv:1612.09265}\ } (\bibinfo {year} {2016})}\BibitemShut {NoStop}%
\bibitem [{\citenamefont {Tukey}(1977)}]{tukey1977exploratory}%
  \BibitemOpen
  \bibfield  {author} {\bibinfo {author} {\bibfnamefont {J.~W.}\ \bibnamefont
  {Tukey}},\ }\href@noop {} {\  (\bibinfo {year} {1977})}\BibitemShut {NoStop}%
\bibitem [{\citenamefont {McGill}, \citenamefont {Tukey},\ and\ \citenamefont
  {Larsen}(1978)}]{mcgill1978variations}%
  \BibitemOpen
  \bibfield  {author} {\bibinfo {author} {\bibfnamefont {R.}~\bibnamefont
  {McGill}}, \bibinfo {author} {\bibfnamefont {J.~W.}\ \bibnamefont {Tukey}}, \
  and\ \bibinfo {author} {\bibfnamefont {W.~A.}\ \bibnamefont {Larsen}},\
  }\href@noop {} {\bibfield  {journal} {\bibinfo  {journal} {The American
  Statistician}\ }\textbf {\bibinfo {volume} {32}},\ \unskip\ \bibinfo {pages}
  {12--16} (\bibinfo {year} {1978})}\BibitemShut {NoStop}%
\bibitem [{\citenamefont {Parzen}(1979)}]{parzen1979nonparametric}%
  \BibitemOpen
  \bibfield  {author} {\bibinfo {author} {\bibfnamefont {E.}~\bibnamefont
  {Parzen}},\ }\href@noop {} {\bibfield  {journal} {\bibinfo  {journal}
  {Journal of the American statistical association}\ }\textbf {\bibinfo
  {volume} {74}},\ \unskip\ \bibinfo {pages} {105--121} (\bibinfo {year}
  {1979})}\BibitemShut {NoStop}%
\bibitem [{\citenamefont {Hyndman}\ and\ \citenamefont
  {Fan}(1996)}]{hyndman1996sample}%
  \BibitemOpen
  \bibfield  {author} {\bibinfo {author} {\bibfnamefont {R.~J.}\ \bibnamefont
  {Hyndman}}\ and\ \bibinfo {author} {\bibfnamefont {Y.}~\bibnamefont {Fan}},\
  }\href@noop {} {\bibfield  {journal} {\bibinfo  {journal} {The American
  Statistician}\ }\textbf {\bibinfo {volume} {50}},\ \unskip\ \bibinfo {pages}
  {361--365} (\bibinfo {year} {1996})}\BibitemShut {NoStop}%
\bibitem [{\citenamefont {Langford}(2006)}]{langford2006quartiles}%
  \BibitemOpen
  \bibfield  {author} {\bibinfo {author} {\bibfnamefont {E.}~\bibnamefont
  {Langford}},\ }\href@noop {} {\bibfield  {journal} {\bibinfo  {journal}
  {Journal of Statistics Education}\ }\textbf {\bibinfo {volume} {14}},\
  \unskip\ \bibinfo {pages} {1--27} (\bibinfo {year} {2006})}\BibitemShut
  {NoStop}%
\bibitem [{\citenamefont {Devore}(2015)}]{devore2015probability}%
  \BibitemOpen
  \bibfield  {author} {\bibinfo {author} {\bibfnamefont {J.~L.}\ \bibnamefont
  {Devore}},\ }\href@noop {} {\emph {\bibinfo {title} {Probability and
  Statistics for Engineering and the Sciences}}}\ (\bibinfo  {publisher}
  {Cengage Learning},\ \bibinfo {year} {2015})\BibitemShut {NoStop}%
\bibitem [{\citenamefont {SEMATHECH}()}]{NISTSEMATHECH}%
  \BibitemOpen
  \bibfield  {author} {\bibinfo {author} {\bibfnamefont {N.}~\bibnamefont
  {SEMATHECH}},\ }\href@noop {} {\bibinfo  {journal}
  {http://www.itl.nist.gov/div898/handbook/prc/section1/prc16.htm}\
  }\BibitemShut {NoStop}%
\bibitem [{\citenamefont {Watkins}, \citenamefont {Scheaffer},\ and\
  \citenamefont {Cobb}(2010)}]{watkins2010statistics}%
  \BibitemOpen
\bibfield  {journal} {  }\bibfield  {author} {\bibinfo {author} {\bibfnamefont
  {A.~E.}\ \bibnamefont {Watkins}}, \bibinfo {author} {\bibfnamefont {R.~L.}\
  \bibnamefont {Scheaffer}}, \ and\ \bibinfo {author} {\bibfnamefont {G.~W.}\
  \bibnamefont {Cobb}},\ }\href@noop {} {\emph {\bibinfo {title} {Statistics:
  from data to decision}}}\ (\bibinfo  {publisher} {John Wiley \& Sons},\
  \bibinfo {year} {2010})\BibitemShut {NoStop}%
\bibitem [{\citenamefont {De~Haan}\ and\ \citenamefont
  {Ferreira}(2007)}]{de2007extreme}%
  \BibitemOpen
  \bibfield  {author} {\bibinfo {author} {\bibfnamefont {L.}~\bibnamefont
  {De~Haan}}\ and\ \bibinfo {author} {\bibfnamefont {A.}~\bibnamefont
  {Ferreira}},\ }\href@noop {} {\emph {\bibinfo {title} {Extreme value theory:
  an introduction}}}\ (\bibinfo  {publisher} {Springer Sci. \& Business
  Media},\ \bibinfo {year} {2007})\BibitemShut {NoStop}%
\bibitem [{\citenamefont {Hill}\ \emph {et~al.}(1975)\citenamefont {Hill} \emph
  {et~al.}}]{hill1975simple}%
  \BibitemOpen
  \bibfield  {author} {\bibinfo {author} {\bibfnamefont {B.~M.}\ \bibnamefont
  {Hill}} \emph {et~al.},\ }\href@noop {} {\bibfield  {journal} {\bibinfo
  {journal} {The annals of statistics}\ }\textbf {\bibinfo {volume} {3}},\
  \unskip\ \bibinfo {pages} {1163--1174} (\bibinfo {year} {1975})}\BibitemShut
  {NoStop}%
\bibitem [{\citenamefont {Jordanova}\ and\ \citenamefont
  {Pancheva}(2012)}]{jordanova2012weak}%
  \BibitemOpen
  \bibfield  {author} {\bibinfo {author} {\bibfnamefont {P.}~\bibnamefont
  {Jordanova}}\ and\ \bibinfo {author} {\bibfnamefont {E.}~\bibnamefont
  {Pancheva}},\ }\href@noop {} {\bibfield  {journal} {\bibinfo  {journal}
  {Comptes rendus de l’acad{\'e}mie bulgare des sciences}\ }\textbf {\bibinfo
  {volume} {65}},\ \unskip\ \bibinfo {pages} {1649--1656} (\bibinfo {year}
  {2012})}\BibitemShut {NoStop}%
\bibitem [{\citenamefont {Fabi{\'a}n}\ and\ \citenamefont
  {Stehl{\i}k}(2009)}]{fabian2009ifas}%
  \BibitemOpen
  \bibfield  {author} {\bibinfo {author} {\bibfnamefont {Z.}~\bibnamefont
  {Fabi{\'a}n}}\ and\ \bibinfo {author} {\bibfnamefont {M.}~\bibnamefont
  {Stehl{\i}k}},\ }\href@noop {} {\  (\bibinfo {year} {2009})}\BibitemShut
  {NoStop}%
\bibitem [{\citenamefont {Embrechts}(1997)}]{EMK}%
  \BibitemOpen
  \bibfield  {author} {\bibinfo {author} {\bibfnamefont {K.~C. M.~T.}\
  \bibnamefont {Embrechts}, \bibfnamefont {P.}},\ }\href@noop {} {\emph
  {\bibinfo {title} {Modelling Extremal Events for Insurance and Finance.}}}\
  (\bibinfo  {publisher} {Springer},\ \bibinfo {year} {1997})\BibitemShut
  {NoStop}%
\end{thebibliography}

\end{document}